# The social dilemma of autonomous vehicles

Jean-François Bonnefon[1], Azim Shariff[2], and Iyad Rahwan[*3]

[1] Toulouse School of Economics, Institute for Advanced Study in Toulouse, Center for Research in Management, University of Toulouse, France

[2] Department of Psychology, University of Oregon, Eugene, OR 97403, USA

[3] The Media Lab, Massachusetts Institute of Technology, Cambridge, MA 02139, USA

Autonomous Vehicles (AVs) should reduce traffic accidents, but they will sometimes have to choose between two evils—for example, running over pedestrians or sacrificing itself and its passenger to save them. Defining the algorithms that will help AVs make these moral decisions is a formidable challenge. We found that participants to six MTurk studies approved of utilitarian AVs (that sacrifice their passengers for the greater good), and would like others to buy them; but they would themselves prefer to ride in AVs that protect their passengers at all costs. They would disapprove of enforcing utilitarian AVs, and would be less willing to buy such a regulated AV. Accordingly, regulating for utilitarian algorithms may paradoxically increase casualties by postponing the adoption of a safer technology.

[*] Correspondence should be addressed to irahwan@mit.edu

The year 2007 saw the completion of the first benchmark test for autonomous driving in realistic urban environments [1, 2]. Since then, Autonomous Vehicles (AVs) such as the Google Car covered thousands of miles of real-road driving [3]. AVs promise world-changing benefits, by increasing traffic efficiency [4], reducing pollution [5], and eliminating up to 90% of traffic accidents [6]. Not all crashes will be avoided, though, and some crashes will require AVs to make difficult ethical decisions, in cases which involve unavoidable harm [7]. For example, the AV may avoid harming several pedestrians by swerving and sacrificing a passer-by, or the AV may be faced with the choice of sacrificing its own passenger to save one or more pedestrians (Fig. 1).

Although these scenarios appear unlikely, even low-probability events are bound to occur with millions of AVs on the road. Moreover, even if these situations were never to arise, programming must nevertheless include decision rules about what to do such in hypothetical situations. Thus these types of decisions need be made well before AVs become a global market. Distributing harm is a decision that is universally considered to fall within the moral domain [8, 9]. Accordingly, the algorithms that control AVs will need to embed moral principles guiding their decisions in situations of unavoidable harm [10]. Manufacturers and regulators will need to accomplish three potentially incompatible objectives: being consistent, not causing public outrage, and not discouraging buyers.

However, pursuing these objectives may lead to moral inconsistencies. Consider for example the case displayed in Fig. 1a, and assume that the most common moral attitude is that the AV should swerve. This would fit a utilitarian moral doctrine [11], according to which the moral course of action is to minimize casualties. But consider then the case displayed in Fig. 1c. The utilitarian course of action, in that situation, would be for the AV to swerve and kill its passenger—but AVs programmed to follow this course of action might discourage buyers who believe their own safety should trump other considerations. Even though such situations may be exceedingly rare, their emotional saliency is likely to give them broad public exposure and a disproportionate weight in individual and public decisions about AVs. To align moral algorithms with human values, we must start a collective discussion about the ethics of AVs, that is, the moral algorithms that we are willing to accept as citizens and to be subjected to as car owners. Here we initiate the data-driven study of driverless car ethics, inspired by the methods of experimental ethics [12].

We conducted six online surveys between June and November 2015. All studies were programmed on Qualtrics survey software and recruited participants (USA residents only) from the Mechanical Turk platform, for a compensation of 25 cents. The experimental ethics literature largely relies on MTurk respondents, with robust results, even though Turk respondents are not necessarily representative of the US population [13,14]. A possible concern with MTurk studies is that some participants may already be familiar with testing materials, when these materials are used by many research groups; but this concern does not apply to our testing materials, which have never been used in a published MTurk study so far.

In all studies, participants provided basic demographic information. Regression analyses (see Supplementary online materials) showed that enthusiasm for self-driving cars was consistently greater for younger, male participants. Accordingly, all subsequent analyses included age and sex as covariates. The last item in every study was an easy question relative to the traffic situation that participants had just considered, e.g., how many pedestrians were on the road. Participants who failed this attention check (typically 10% of the sample) were discarded from subsequent analyses.

Detailed statistical results for all studies are provided in Supplementary online materials (Tables S1-S8). Overall, participants strongly agreed that it would be more moral for AVs to sacrifice their own passengers, when this sacrifice would save a greater number of lives overall.

In Study 1 ($n = 182$), 76% of participants thought that it would be more moral for AVs to sacrifice one passenger, rather than kill ten pedestrians (with a 95% confidence interval of 69—82). These same participants were later asked to rate which was the most moral way to program AVs, on a scale from 0 (protect the passenger at all costs) to 100 (minimize the number of casualties). They overwhelmingly expressed a moral preference for utilitarian AVs programmed to minimize the number of casualties (median = 85, Fig. 2a). Participants were not quite so sure that AVs would be programmed that way though (67% thought so, with a median rating of 70). Thus, participants were not worried about AVs being too utilitarian, as often portrayed in science-fiction. If anything, they imagined future AVs as being *less* utilitarian than they should be.

In Study 2 (*n* = 451), participants were presented with dilemmas that varied the number of pedestrians' lives that could be saved, from 1 to 100. Participants did not think that AVs should sacrifice their passenger when only one pedestrian could be saved (with an average approval rate of 23%), but their moral approval increased with the number of lives that could be saved ($p < 0.001$) up to approval rates consistent with the 76% observed in Study 1 (Fig. 2b),.

Participants' approval of passenger sacrifice was even robust to treatments in which they had to imagine themselves and another person in the AV, a family member in particular (Study 3, *n* = 259). Imagining that a family member was in the AV negatively impacted the morality of the sacrifice, as compared to imagining oneself alone in the AV ($p = 0.003$), but even in that strongly aversive situation, the morality of the sacrifice was still rated above the midpoint of the scale, with a 95% confidence interval of 54–66 (Fig. 3a).

But this study presents the first hint of a social dilemma. On a scale of 1–100, respondents were asked to indicate how likely they would be to buy an AV programmed to minimize casualties (which would, in these circumstances, sacrifice them and their co-rider family member), and how likely they would be to buy an AV programmed to prioritize protecting its passengers, even if it meant killing 10 or 20 pedestrians. Although the reported likelihood of buying an AV was low even for the self-protective option (median = 50), respondents indicated a significantly lower likelihood ($p < .001$) of buying the AV when they imagined the situation in which they and their family member would be sacrificed for the greater good (median = 19). In other words, even though participants still agreed that utilitarian AVs were the most moral, they preferred the self-protective model for themselves.

Study 4 (*n* = 267) offers another demonstration of this phenomenon. Participants were given 100 points to allocate between different types of algorithms, once to indicate how moral the algorithms were, once to indicate how comfortable they were for other AVs to be programmed that way, and once for how likely they would be to buy an AV programmed that way. For one of the algorithms, the AV would always swerve when it was about to run over people on the road. Fig 3b shows the points allocated to the AV equipped with this algorithm, in three situations: when it swerved into a pedestrian to save 10 people, when it killed its own passenger to save 10 people, and when it swerved into a pedestrian to save just one other pedestrian. The algorithm that swerved into one to save 10 always received many points, and the algorithm that swerved into one to save one always received few points. The algorithm that would kill its passenger to save 10 presented a hybrid profile. Just as the high-valued algorithm, it received high marks for morality (median budget share = 50), and was considered a good algorithm for other people to have (median budget share = 50). But in terms of purchase intention, it received significantly fewer points than the high-valued algorithm ($p < 0.001$), and was in fact closer to the low-valued algorithms (median budget share = 33). Once more, it appears that people praise utilitarian, self-sacrificing AVs, and welcome them on the road, without actually wanting to buy one for themselves.

This is the classic signature of a social dilemma, in which everyone has a temptation to free-ride instead of adopting the behavior that would lead to the best global outcome. One typical solution in this case is for regulators to enforce the behavior leading to the best global outcome; indeed, there are many other examples of tradeoff by people and governments that involve harm [15-17]. For example, some citizens object to regulations that require to immunize children before they start school, minimizing the perceived risk of harm to their child, while increasing the risk to others; and recognition of the threats of environmental degradation have prompted government regulations aimed at curtailing harmful behaviors for the greater good. But would people approve of government regulations imposing utilitarian algorithms in AVs, and would they be more likely to buy AVs under such a regulation?

In Study 5 ($n = 376$), we asked participants about their attitudes towards legally enforcing utilitarian sacrifices. Participants considered scenarios in which either a human driver or a control algorithm had an opportunity to self-sacrifice in order to save 1 or 10 pedestrians (Fig 3c). As usual, the perceived morality of the sacrifice was high, and about the same whether the sacrifice was performed by a human or by an algorithm (median = 70). When asked whether they would agree to see such moral sacrifices legally enforced, their agreement was higher for algorithms than for human drivers ($p < 0.002$), but the average agreement still remained below the midpoint of the 0–100 scale in each scenario. Agreement was highest in the scenario where algorithms saved 10 lives, with a 95% confidence interval of 33–46.

Finally, in Study 6 ($n = 393$), we asked participants specifically about their likelihood of purchasing the AVs whose algorithms had been regulated by the government. Participants were presented with scenarios in which they were either riding alone, with an unspecified family member, or with their child. As in the previous studies the scenarios depicted a situation in which the algorithm that controlled the AV could sacrifice its passengers to minimize casualties on the road. Participants indicated whether it was the duty of the government to take regulations that would minimize the casualties in such circumstances, whether they would consider the purchase of an AV under such a regulation, and whether they would consider purchasing an AV under no such regulation. As shown in Fig 3d, people were reluctant to accept governmental regulation of utilitarian AVs. Even in the most favorable condition, when they were only imagining themselves being sacrificed to save 10 pedestrians, the 95% confidence interval for whether people thought it was appropriate for the government to regulate this sacrifice was only 36–48. Finally and importantly, participants were vastly less likely to consider purchasing an AV with such regulation than without ($p < 0.001$). The median expressed likelihood of purchasing an unregulated AV was 59, compared to 21 for purchasing a regulated AV. This is a huge gap from a statistical perspective, but it must be understood as reflecting the state of public sentiment at the very beginning of a new public issue, and thus not guaranteed to persist.

Three groups may be able to decide how AVs handle ethical dilemmas: the consumers who buy the AVs; the manufacturers that must program the AVs in the first place; and the government which may regulate the kind of programming that manufacturers can offer and consumers can choose from. While manufacturers may engage in advertising and lobbying to influence the preferences of consumers and the regulations that the governments may pass, a critical collective problem consists of deciding whether governments *should* regulate the moral algorithms that manufacturers offer to consumers.

Our findings suggests that regulation may be necessary, but at the same time counter-productive. Moral algorithms for AVs create a social dilemma [18, 19]. Even though people seem to agree that everyone would be better off if AVs were utilitarian (in the sense of minimizing the number of casualties on the road), they all have a personal incentive to ride in AVs that will protect them at all costs. Accordingly, if both self-protective and utilitarian AVs were allowed on the market, few would be willing to ride in utilitarian AVs, even though they would prefer others to do so. Regulation may provide a solution to this problem, but it faces two difficulties. First, most people seem to disapprove of a regulation that would enforce utilitarian AVs. Second and worse, our results suggest that such a regulation could significantly delay the adoption of AVs— which means that the lives saved by making AVs utilitarian may be outnumbered by the deaths caused by delaying the adoption of the AVs altogether, with their other safety enhancing technologies. This is a challenge that should be on the mind of carmakers and regulators alike.

Moral algorithms for AVs will need to tackle more intricate decisions than that we have considered in our surveys. For example, our scenarios did not feature any uncertainty about decision outcomes, but a collective discussion about moral algorithms will have to tackle the concepts of expected risk, expected value, and blame assignment. Is it acceptable for an AV to avoid a motorcycle by swerving into a wall, considering that the probability of survival is greater for the passenger of the AV, than for the rider of the motorcycle? Should AVs take the ages of the passengers and pedestrians into account [20]? If a manufacturer offers different versions of its moral algorithm, and a buyer knowingly chose one of them, is the buyer to blame for the harmful consequences of the algorithm's decisions? Such liability considerations will need to accompany existing discussions of regulation [21], and we hope that psychological studies inspired by our own will be able to inform this discussion.

Figuring out how to build ethical autonomous machines is one of the thorniest challenges in artificial intelligence today [22]. As we are about to endow millions of vehicles with autonomy, taking algorithmic morality seriously has never been more urgent. Our data-driven approach highlights how the field of experimental ethics can give us key insights into the moral, cultural and legal standards that people expect from autonomous driving algorithms. For the time being, there seems to be no easy way to design algorithms that would reconcile moral values and personal self-interest, let alone across different cultures with different moral attitudes to life-life tradeoffs [23]—but public opinion and social pressure may very well shift as this conversation progresses.

# References

[1]  Montemerlo B et al. Junior: The Stanford entry in the urban challenge. *Journal of Field Robotics*, 25:569–597, 2008.

[2]  Urmson C et al. Autonomous driving in urban environments: Boss and the urban challenge. *Journal of Field Robotics*, 25:425–266, 2008.

[3]  Waldrop MM. Autonomous vehicles: No drivers required. *Nature*, 518:20–23, 2015.

[4]  Van Arem B, Van Driel CJ, Visser R. The impact of cooperative adaptive cruise control on traffic-flow characteristics. *IEEE Transactions on Intelligent Transportation Systems*, 7:429–436, 2006.

[5]  Spieser K et al. Toward a systematic approach to the design and evaluation of automated mobility-on-demand systems: A case study in Singapore. In Meyer G, Beiker S, editor, *Road Vehicle Automation*, pages 229–245. Springer, 2014.

[6]  Gao P, Hensley R, Zielke A. A roadmap to the future for the auto industry, 2014.

[7]  Goodall NJ. Machine ethics and automated vehicles. In Meyer G, Beiker S, editor, *Road Vehicle Automation*, pages 93–102. Springer, 2014.

[8]  Gray K, Waytz A, and Young L. The moral dyad: A fundamental template unifying moral judgment. *Psychological Inquiry*, 23:206–215, 2012.

[9]  Haidt J. *The righteous mind: Why good people are divided by politics and religion*.

Pantheon Books, 2012.

[10]  Wallach W, Allen C. *Moral machines: Teaching robots right from wrong*. Oxford University Press, 2008.

[11]  Rosen F. *Classical utilitarianism from Hume to Mill*. Routledge, 2005.

[12]  Greene JD. *Moral tribes: Emotion, reason and the gap between us and them*. Atlantic Books, 2014.

[13] Côté S, Piff PK, Willer R. For whom do the ends justify the means? Social class and utilitarian moral judgment. *Journal of Personality and Social Psychology,* 104: 490-503., 2013

[14] Everett JAC, Pizarro DA, Crockett MJ. Inference of trustworthiness from intuitive moral judgments. *Journal of Experimental Psychology: General*, 2016.

[15] Kass NE. An ethics framework for public health. *American Journal of Public Health*, 91: 1776–1782, 2001.

[16] Sunstein CR, Vermeule A. Is capital punishment morally required? Acts omissions, and life-life tradeoffs. *Stanford Law Review,* 58: 703–750, 2005.

[17] Dietz T, Ostrom E, Stern PC. The struggle to govern the commons. *Science,* 302: 1907-1912, 2003.

[18]  Dawes RM. Social dilemmas. *Annual Review of Psycholology*, 31:169–193, 1980.

[19]  Van Lange PAM, Joireman J, Parks CD, Craig D, Van Dijk E. The psychology of social dilemmas: A review. *Organizational Behavior and Human Decision Processes*, 120:125–141, 2013.

[20] Posner EA, Sunstein CR. Dollars and death. *The University of Chicago Law Review,* 72: 537–598, 2005.

[21] Vladeck DC. Machines without principals: Liability rules and Artificial Intelligence. *Washington Law Review,* 89: 117-150, 2014

[21] Deng B. Machine ethics: The robot's dilemma. *Nature*, 523:24–26, 2015.

[22] Gold N, Colman AM, Pulford, BD. Cultural differences in response to real-life and hypothetical trolley problems. *Judgment and Decision Making*, 9: 65–76, 2014.

Bonnefon gratefully acknowledges support through the ANR-Labex IAST. This research was supported by internal funds from the University of Oregon to Shariff. Rahwan is grateful for financial support from Reid Hoffman. Data files have been uploaded as Supplementary Online Material.

Figure 1: Three traffic situations involving imminent unavoidable harm. The car must decide between (a) killing several pedestrians or one passer by, (b) killing one pedestrian or killing its own passenger, (c) killing several pedestrians or killing its own passenger.

Figure 2: Considering the greater good vs the life of the passenger. In studies 1 and 2, when asked which would be the most moral way to program AVs, participants expressed a preference for AVs programmed to kill their passenger for the greater good. This preference was strong as soon as 5 lives or more could be saved (left panel shows detailed results for 10 lives). On average, participants were more confident that AVs *should* pursue the greater good than they were confident that AVs *would* be programmed to do so. Boxes show the 95% confidence interval of the mean.

Fig. 3: Towards regulation and purchase (studies 3-6). Boxes show the 95% confidence interval of the mean. In all studies, participants expressed a moral preference for AVs sacrificing their passengers to save a greater number of pedestrians. This moral preference was robust to situations in which participants imagined themselves in the AV in the company of a coworker, a family member, or their own child. However, (a and b) participants did not ex- press a comparable preference for buying utilitarian AVs, especially when they thought of family members riding in the car, and (c and d) they disapproved of regulations enforcing utilitarian algorithms for AVs, and indicated that they would be less likely to purchase an AV under such a regulation.

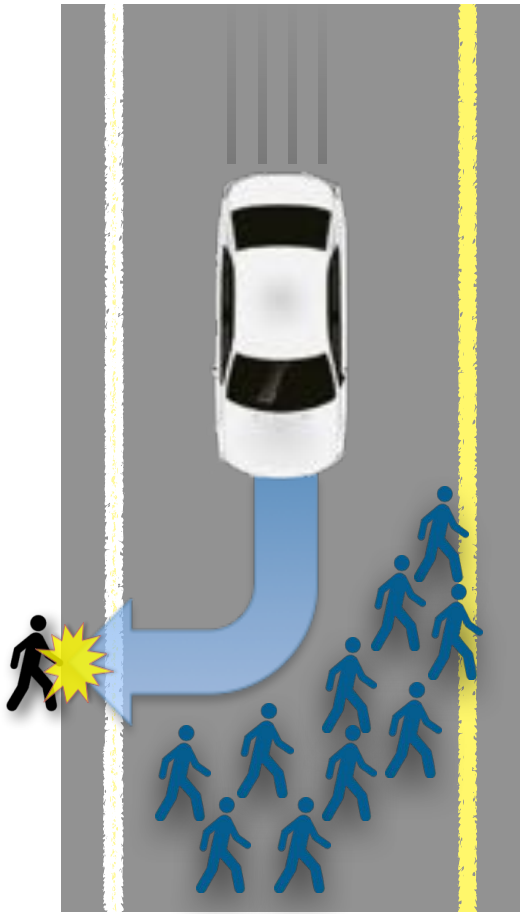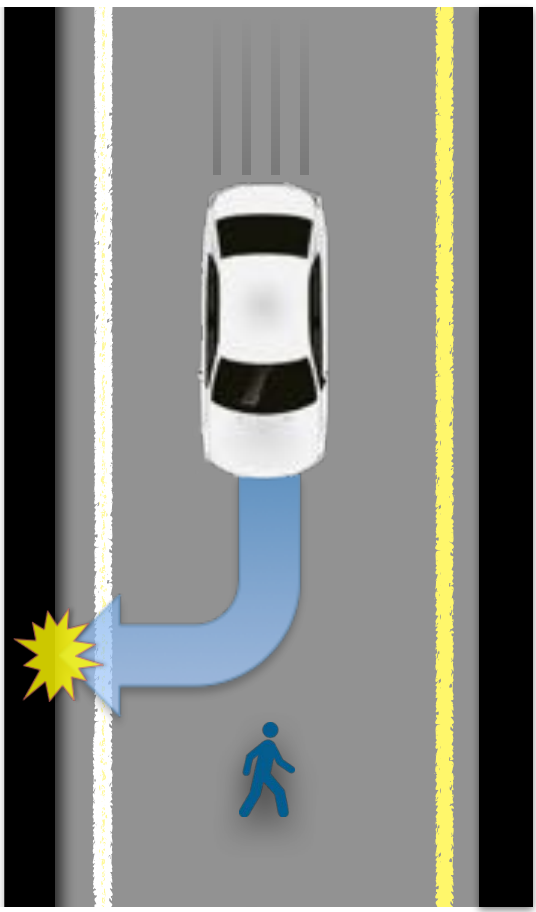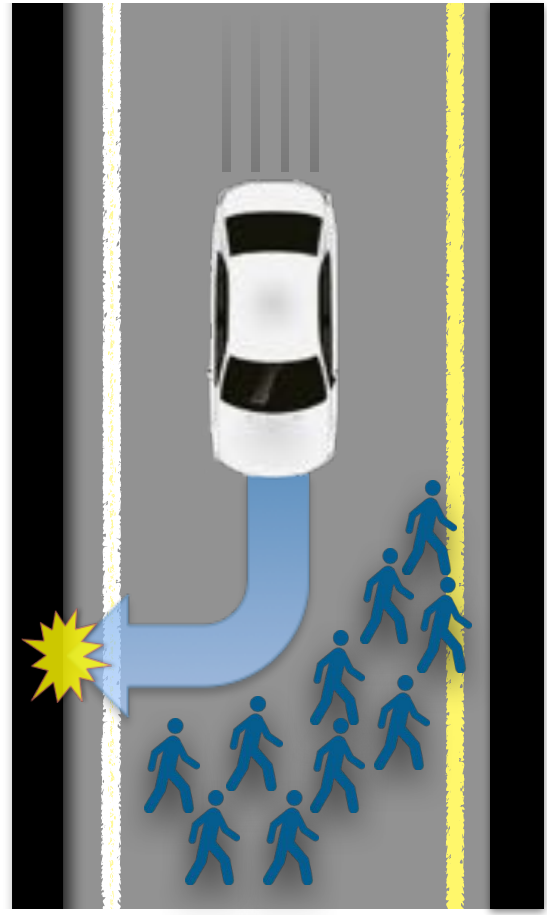

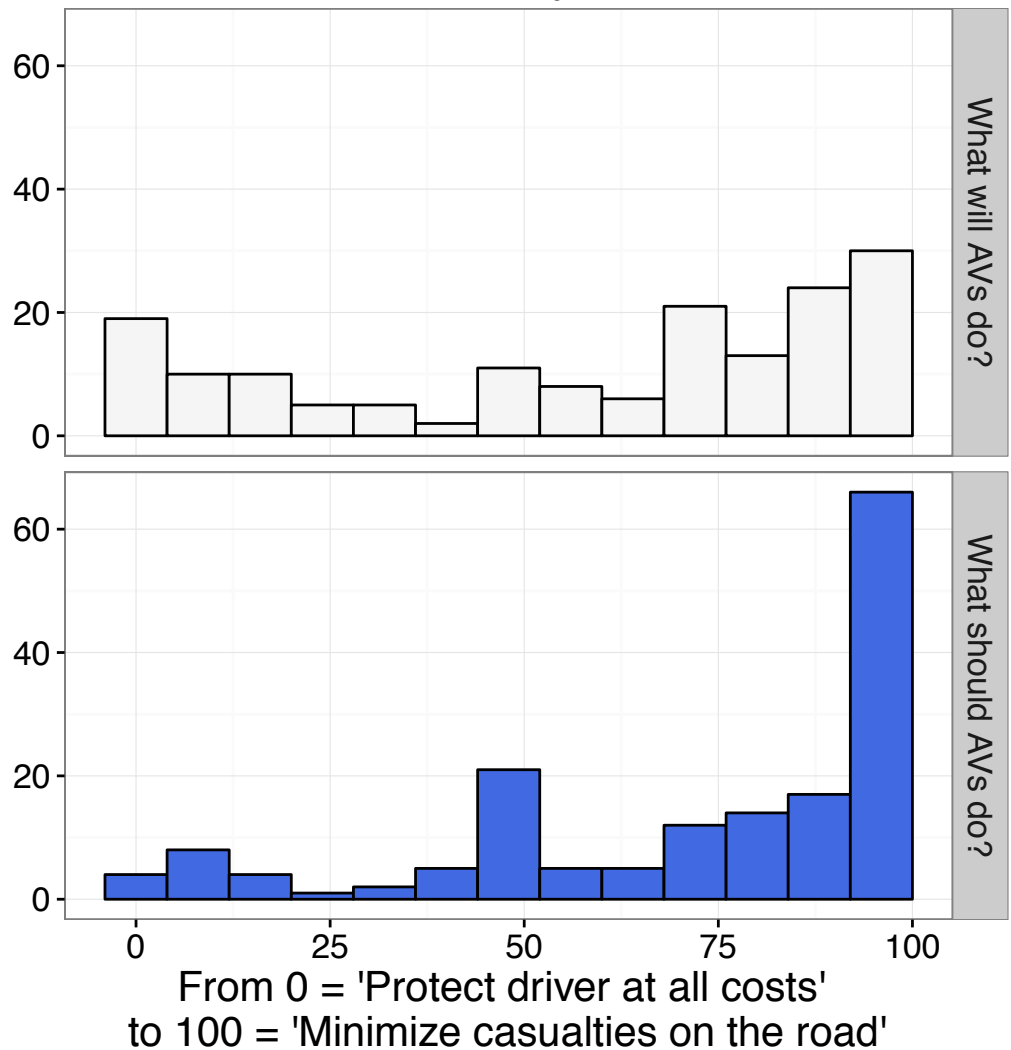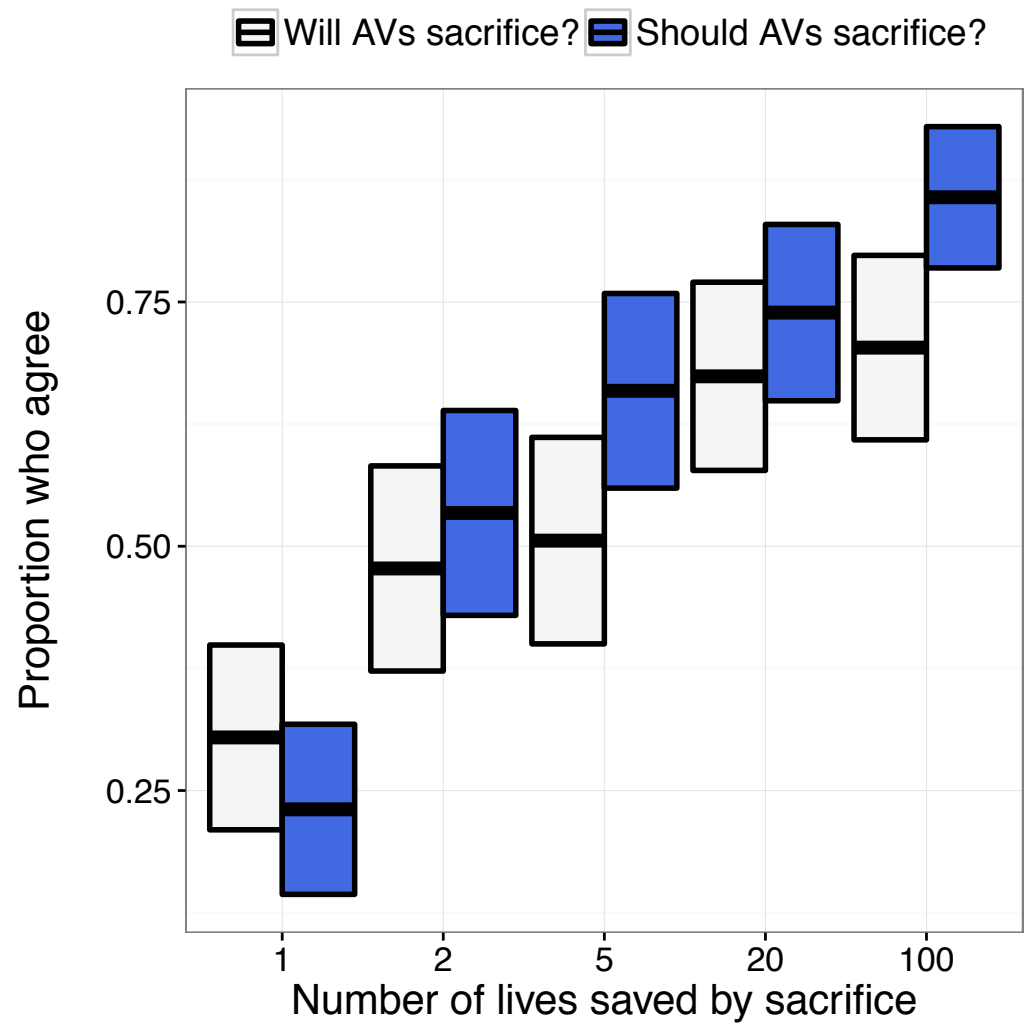

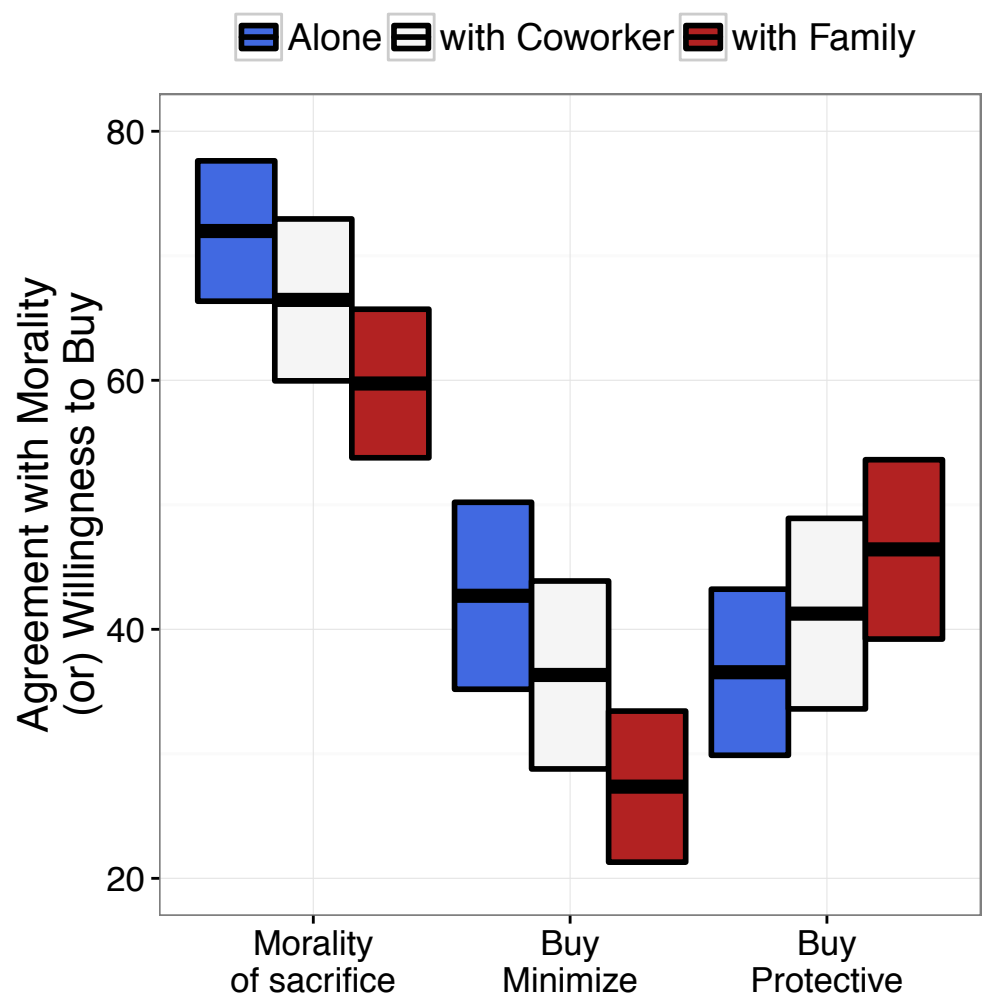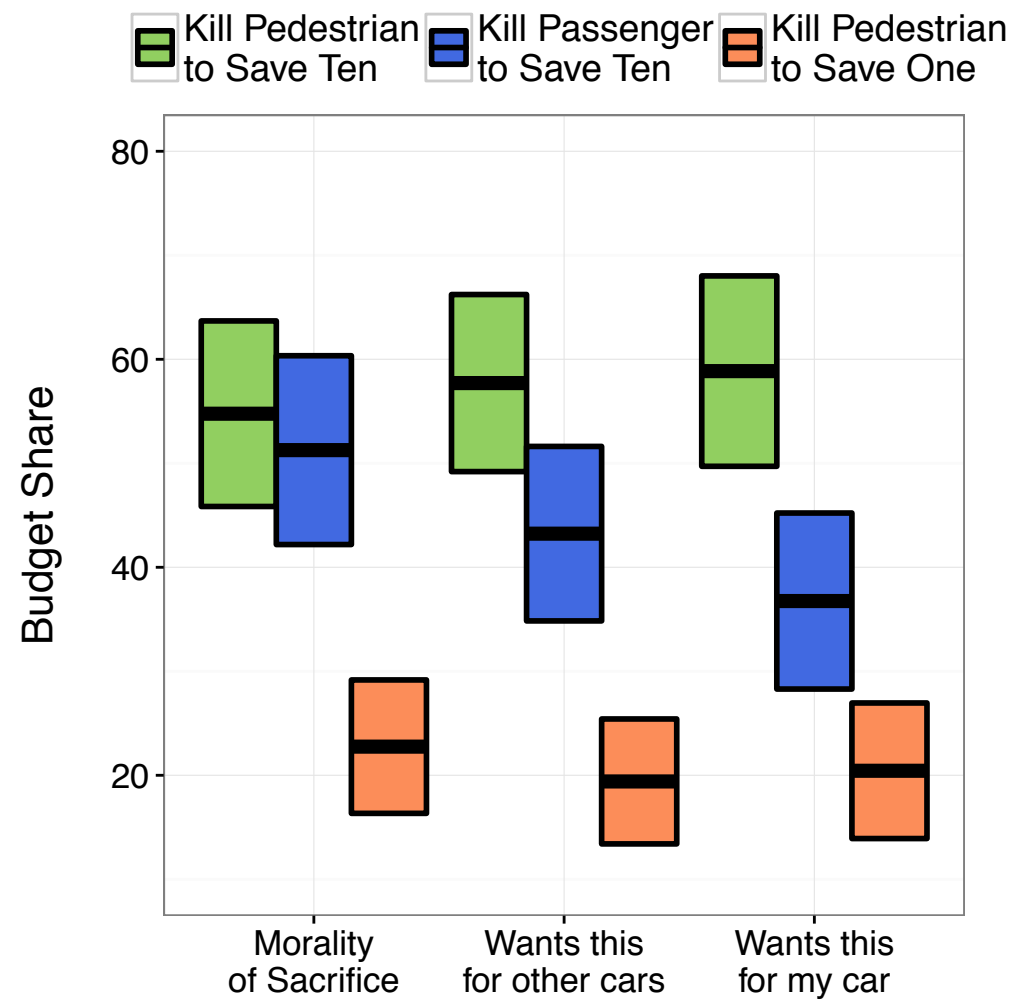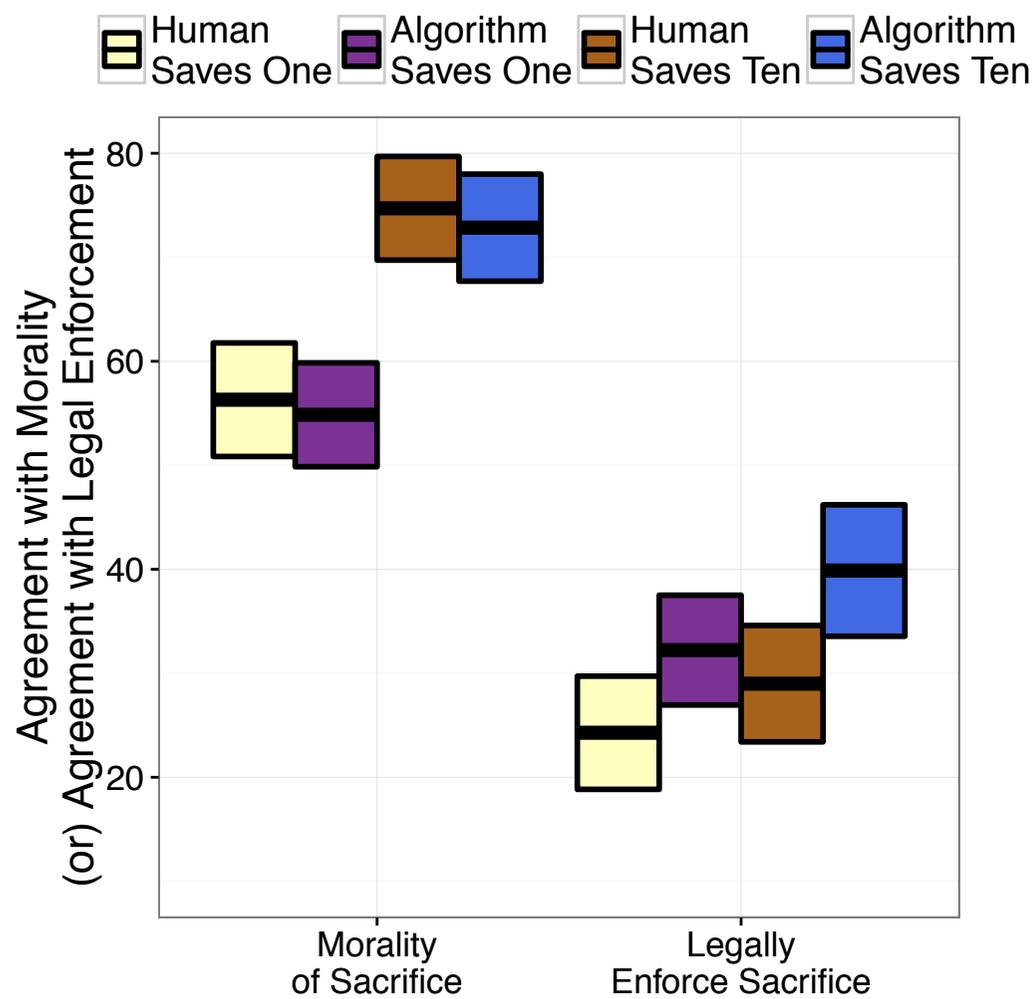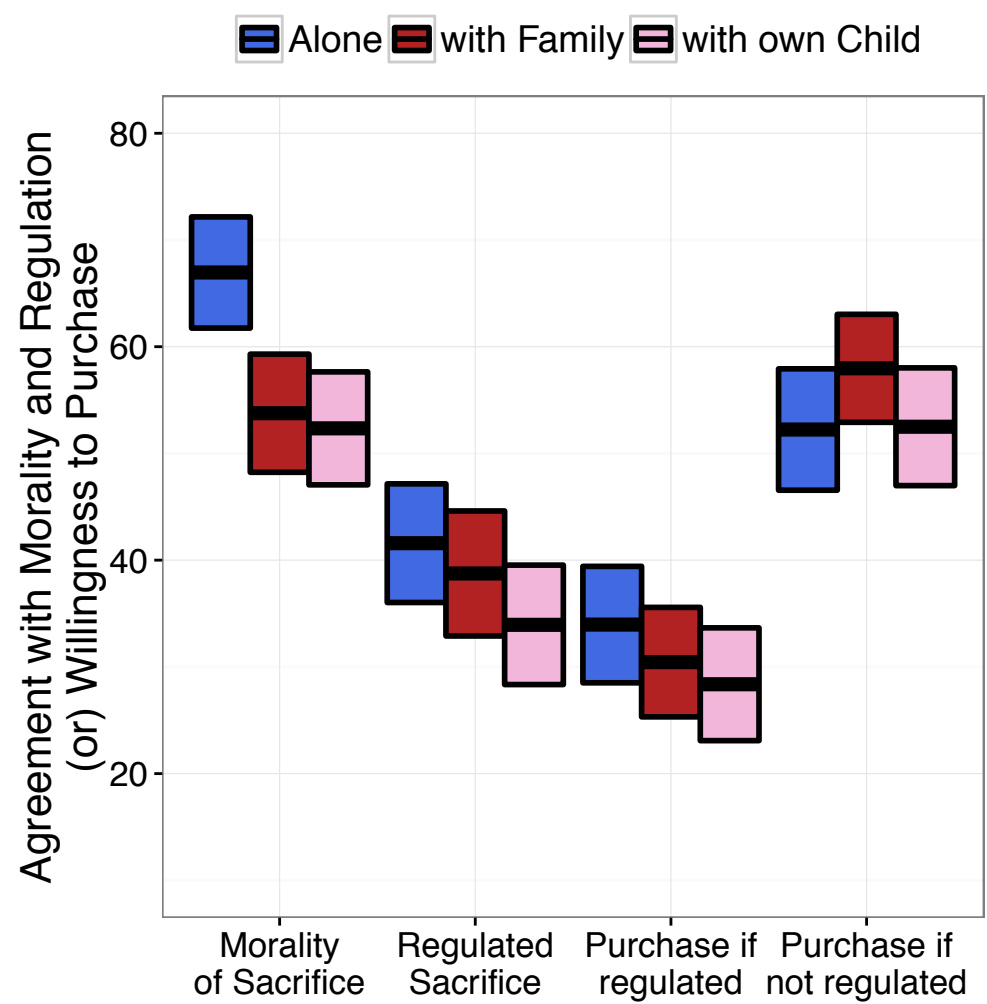